\documentclass[12pt]{article}
\usepackage{graphicx}

\newcommand\pubdate{\today}

\newcommand\pubnumber{TUM-HEP-813/11}

\def\Title#1{\begin{center} {\Large #1 } \end{center}}
\def\Author#1{\begin{center}{ \sc #1} \end{center}}
\def\Address#1{\begin{center}{ \it #1} \end{center}}

\newcommand\pubblock{\rightline{\begin{tabular}{l} \pubnumber\\
         \pubdate  \end{tabular}}}
\newenvironment{Abstract}{\begin{center}{\bf Abstract}\end{center} \bigskip \begin{quotation}  }{\end{quotation}}
\newenvironment{Presented}{\begin{quotation} \begin{center} 
             PRESENTED AT\end{center}\bigskip 
      \begin{center}\begin{large}}{\end{large}\end{center} \end{quotation}}
\def\Acknowledgements{\bigskip  \bigskip \begin{center} \begin{large}
             \bf ACKNOWLEDGEMENTS \end{large}\end{center}}

\def\beq{\begin{equation}}
\def\eeq#1{\label{#1}\end{equation}}
\def\eeqn{\end{equation}}

\def\beqa{\begin{eqnarray}}
\def\eeqa#1{\label{#1}\end{eqnarray}}
\def\eeqan{\end{eqnarray}}

\let\bar=\overbar

\def\Dslash{\not{\hbox{\kern-4pt $D$}}}
\def\dslash{\not{\hbox{\kern-2pt $\del$}}}

\def\msb{{\bar{\ssstyle M \kern -1pt S}}}

\begin{document}
\begin{titlepage}
\pubblock

\vfill


\Title{Theoretical status of the CKM Matrix}
\vfill
\Author{Alexander Lenz}  
\Address{Physik Department, Technische Universit{\"a}t M{\"u}nchen, D-85748 Garching, Germany}
\vfill


\begin{Abstract}
In this talk I review the current status of the CKM matrix. 
A special emphasis is also given to several discrepancies between 
experiments and the standard model  at the level of about 
three standard deviations. Recent results that appeared after FPCP2011
are also included in the discussion.
\end{Abstract}

\vfill

\begin{Presented}
The Ninth International Conference on\\
Flavor Physics and CP Violation\\
(FPCP 2011)\\
Maale Hachamisha, Israel,  May 23--27, 2011
\end{Presented}
\vfill

\end{titlepage}
\def\thefootnote{\fnsymbol{footnote}}
\setcounter{footnote}{0}
%


\section{Introduction}

The Cabibbo-Kobayashi-Maskawa-matrix \cite{Cabibbo:1963yz,Kobayashi:1973fv}
describes flavor transitions in the quark sector. Its elements
have been investigated in the last years
in detail in particular by the 
PDG \cite{Nakamura:2010zzi}, 
the HFAG \cite{Asner:2010qj} and the collaborations 
CKMfitter \cite{Charles:2004jd} and UTfit \cite{Ciuchini:2000de}.
A recent fit \cite{Lenz:2010gu} gives e.g. the following values for the 
CKM matrix
	\begin{eqnarray}
	V_{CKM} & = &
	\left( \begin{array}{ccc} 
0.97426 \pm 0.00030           & 0.22545 \pm 0.00095 & 0.00356 \pm 0.00020 \\
0.22529 \pm 0.00077           & 0.97341 \pm 0.00021 & 0.04508^{+0.00075}_{-0.00528}\\
0.00861^{+0.00021}_{-0.00037} & 0.04068 \pm 0.00138 & 0.999135^{+0.000057}_{-0.000018}\\
	\end{array} \right)
\label{CKMfit} \; .
       	\end{eqnarray}
The precision of the individual elements is quite impressive.
Also the fit of the so-called unitarity triangle shows a good overall
consistency (figure from \cite{Lenz:2010gu}).

\begin{center}
\includegraphics[width=0.65\textwidth,angle=0]{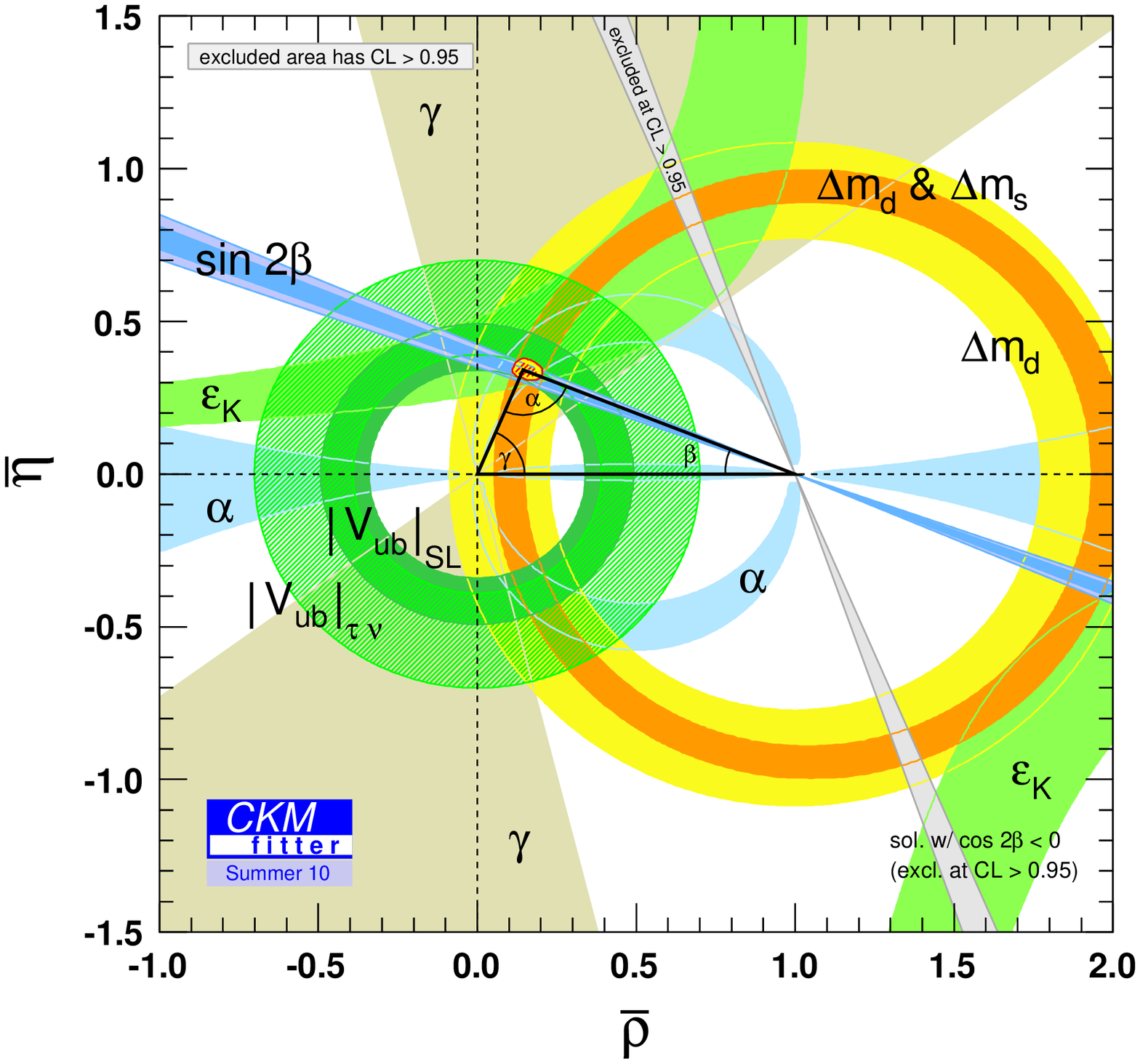}
\end{center}

Similar recent results can be found e.g. in 
\cite{Bevan:2010gi,Lunghi:2010gv,Laiho:2011nz}.
Due to the success of the CKM-picture, Kobayashi and Maskawa were awarded 
in 2008 with the nobel price.
\\
In Section 2 we investigate a little more in detail the determination of 
the individual CKM-elements, while we discuss some hints for deviations
from the standard model in Section 3. In Section 4 we conclude and give 
an outlook. We end with a {\it What to do list} in Section 5.

\section{Status of the individual CKM elements}

In this section we review very briefly the current status of our 
knowledge about the values of the CKM elements. More detailed reviews can 
be found e.g. in \cite{Nakamura:2010zzi,Asner:2010qj,Lenz:2010gu}.

\subsection{First row of the CKM Matrix}

The CKM element $V_{ud}$ is determined in the nuclear $\beta$-decay,
in the neutron $\beta$-decay and in the pion $\beta$-decay 
\cite{Nakamura:2010zzi,Hardy:2008gy}.
Data for the CKM element $V_{us}$ stem from  $K_{l3}$-decays,  
hadronic $\tau$ decays and  semileptonic hyperon decays 
\cite{Nakamura:2010zzi,Antonelli:2010yf,Boyle:2010bh}.
\begin{equation}
\begin{array}{lclcl}
|V_{ud}|                  & = & 0.97425 & \pm & 0.00022 \; ,
\\
|V_{us}|                  & = & 0.2254  & \pm & 0.0013  \; .
\end{array}
\end{equation}
Both elements are quite precisely known, while we have  different
values of $V_{ub}$, depending on the extraction method:
exclusive $B$-decays,
inclusive $B$-decays
(see e.g. the article of Kowalewski and Mannel in \cite{Nakamura:2010zzi} 
for a list of references or \cite{Asner:2010qj}), from $B \to \tau \nu$  
and from a global fit \cite{Lenz:2010gu,Bevan:2010gi}.
\begin{equation}
\begin{array}{lclcl}
|V_{ub}|^{\rm Exclusive}  & = & 0.00351 & \pm & 0.00047 \; ,
\\
|V_{ub}|^{\rm Inclusive}  & = & 0.00432 & \pm & 0.00027 \; ,
\\
|V_{ub}|^{B \to \tau \nu} & = & 0.00510 & \pm & 0.00059 \; ,
\\
|V_{ub}|^{\rm Global Fit} & = & 0.00356  & \pm & 0.00020 \; .
\end{array}
\end{equation}
The third value is taken from \cite{Nierste:2011na}, the rest of the values is
from \cite{Lenz:2010gu}.
Concerning the numerical value of $V_{ub}$ serveral comments are appropriate:
In particular the last two numbers differ quite sizeably, while
the different values for the inclusive and exclusive extraction
might hint to the fact that hadronic uncertainties (e.g. lattice, LCSR)
are underestimated, see also \cite{Bernlochner}. There is also a new physics explanation for this 
discrepancy. Right-handed currents could lead to a deviation of
the exclusive from the inclusive determination 
\cite{Crivellin:2009sd,Buras:2010pz}.
New Physics in $B_d$-mixing might also enhance the global fit value of $V_{ub}$
and therefore reduce the discrepancy with the value extracted from
$B \to \tau \nu$.
Due to these problems with $V_{ub}$ Soni and Lunghi 
(see e.g. \cite{Lunghi:2010gv}) 
suggested not to use $V_{ub}$ in the global fit.
\\
Finally we would like to mention that, $V_{ub}$ is actually of order 
$\lambda^4$ and not of order $\lambda^3$ in the Wolfenstein parameter 
$\lambda \approx 0.2254$ \cite{Wolfenstein:1983yz}
\begin{equation}
0.00356 = (0.2254)^{3.79}\; .
\end{equation}
With all the values of the CKM elements from the first row one
can test the unitarity of the CKM matrix
\begin{eqnarray}
\sqrt{1-V_{ud}^2-V_{us}^2} &= &   0.00564^{+0.02669}_{-0.00564} \; .
\end{eqnarray}
A nice way to study the bounds on the unitarity of the CKM matrix is
to investigate an extension of the standard model with an hypothetical fourth 
generation of fermions. If one assumes that $V_{CKM4}$ is unitary, one
gets the result that $V_{ub'}$ can still be larger than $V_{ub}$ (see e.g. 
\cite{Eberhardt:2010bm,Bobrowski:2009ng,Das:2010fh,Alok:2010zj,Nandi:2010zx,Buras:2010pi}
for some recent studies and also \cite{Ivanov})
\begin{equation}V_{ub'} < 0.04 \; .
\end{equation}
Despite the impressive accuracy of the extracted values of the first row of the
CKM matrix, it is still desirable to reduce the error of $V_{us}$ further
and to clarify the discrepancies in $V_{ub}$. Currently it is still not excluded 
that there exists a value of $V_{ub'}$, which is larger than $V_{ub}$.

\subsection{Second row of the CKM Matrix}

$V_{cd}$ is measured \cite{Nakamura:2010zzi} in
semileptonic charm decays $D \to \pi l \nu$ and in
charm production in neutrino interactions.
$V_{cs}$ is determined \cite{Nakamura:2010zzi} 
in neutrino scattering, on-shell $W$ decays
and in semi-leptonic charm decays. 
$V_{cb}$ is obtained from inclusive $B \to X_c l \nu$
decays and from exclusive $ B \to D^{(*)}$ transitions
(see e.g. the article of Kowalewski and Mannel in \cite{Nakamura:2010zzi} 
for a list of references or \cite{Asner:2010qj}).
\begin{equation}
\begin{array}{lclcl}
|V_{cd}|                  & = & 0.230   & \pm & 0.011 \; ,
\\
|V_{cs}|                  & = & 1.023   & \pm & 0.036 \; ,
\\
|V_{cb}|^{\rm Exclusive}  & = & 0.03885 & \pm & 0.00047 \; ,
\\
|V_{cb}|^{\rm Inclusive}  & = & 0.04115 & \pm & 0.00027 \; .
\\
\end{array}
\end{equation}
Here the uncertainties are considerably larger than in the first row and again the
inclusive determination of $V_{cb}$ yields larger values than the inclusive one.
Sometimes avarages are used for $V_{cb}$, e.g.
\begin{equation}
       |V_{cb}| = \left\{ \begin{array}{ll}
                           (40.6 \pm 1.3) \cdot 10^{-3} & \cite{Nakamura:2010zzi} \; ,
                           \\ 
                           (40.89 \pm 38 \pm 0.59) \cdot 10^{-3} & \cite{Lenz:2010gu} \; .
                           \end{array}
                          \right. 
      \end{equation}
To test the accuracy of the second row
we again investigate a hypothetical 4th generation of fermions
and assume that $V_{CKM4}$ is unitary. One finds 
\cite{Eberhardt:2010bm,Bobrowski:2009ng,Das:2010fh,Alok:2010zj,Nandi:2010zx,Buras:2010pi}
that $V_{cb'}$ can still be considerably larger than $V_{cb}$
\begin{equation}
V_{cb'} < 0.15 \; .
\end{equation}
This is almost the size of the Wolfenstein parameter $\lambda$!
So clearly an improvement in the determination of the CKM elements of the second row
is mandatory.

\subsection{Third row of the CKM Matrix}

\begin{figure}
\includegraphics[width=0.8\textwidth,angle=0]{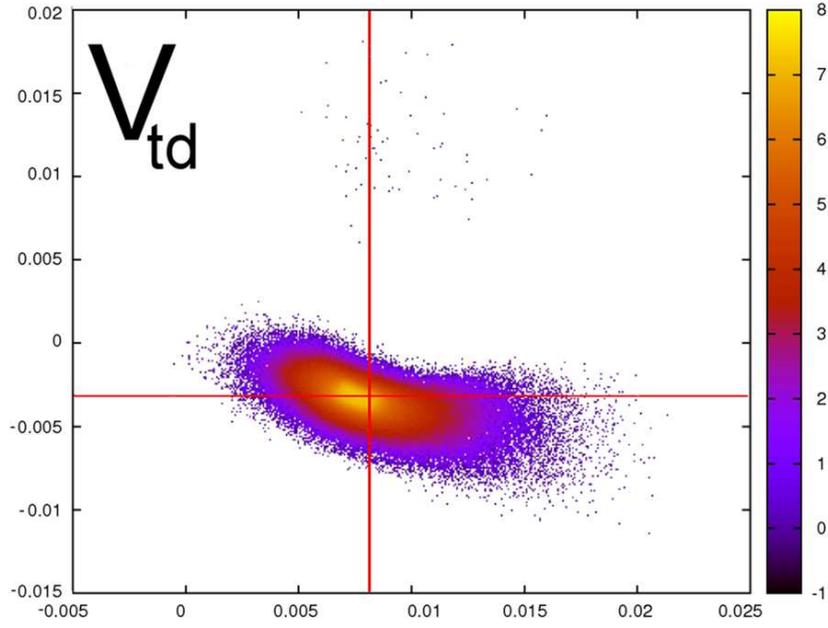}
\caption{Allowed value of $V_{td}$ in a complex plane. The value from
Eq.(\ref{CKMfit}) is denoted by the red lines.}
\label{Vtd}
\end{figure}

\begin{figure}
\includegraphics[width=0.8\textwidth,angle=0]{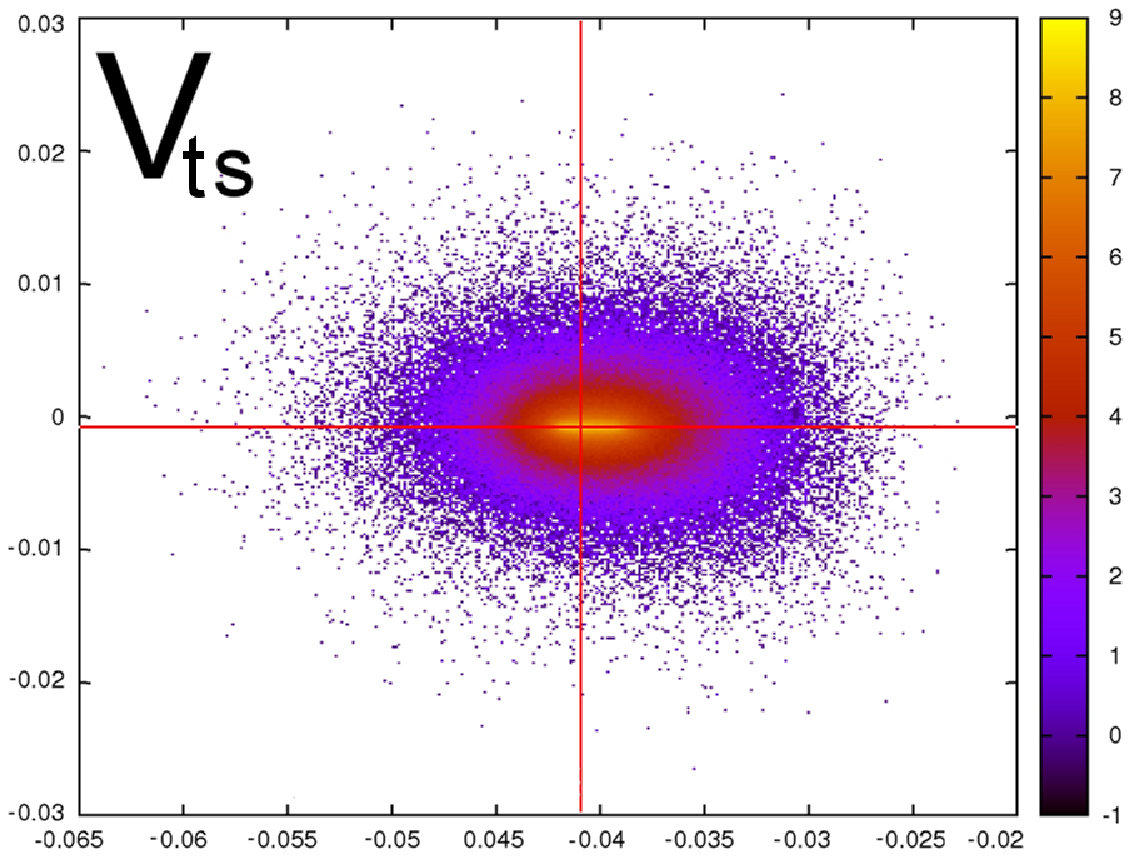}
\caption{Allowed value of $V_{ts}$ in a complex plane. The value from
Eq.(\ref{CKMfit}) is denoted by the red lines.}
\label{Vts}
\end{figure}

\begin{figure}
\includegraphics[width=0.6\textwidth,angle=0]{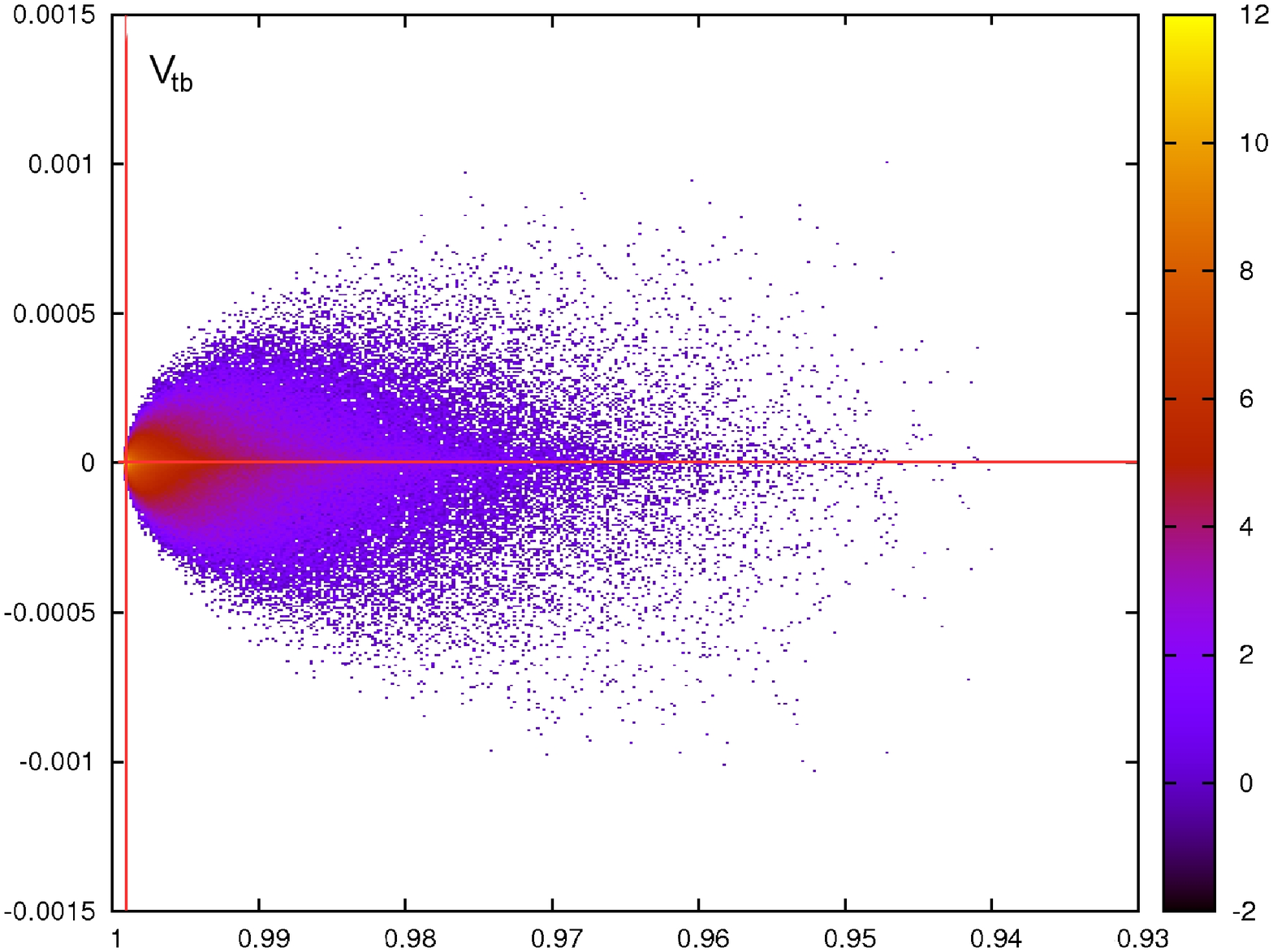}
\caption{Allowed value of $V_{tb}$ in a complex plane. The value from
Eq.(\ref{CKMfit}) is denoted by the red lines.}
\label{Vtb}
\end{figure}

Except for $V_{tb}$ we do not have any direct information about the CKM
elements of the third row. Single top production at the Tevatron
\cite{single-top}
\footnote{After the conference D0 published \cite{Abazov:2011zk} a measurement 
of the ratio ($R= |V_{tb}|^2/(|V_{td}|^2|+V_{ts}|^2+|V_{tb}|^2)= 0.90 \pm 0.04$, 
that gives a more tight bound on $V_{tb}$.}
gives
\begin{equation}
V_{tb} = 0.88 \pm 0.07 \; .
\end{equation}
The precise values for $V_{td}$, $V_{ts}$ and
$V_{tb}$ from Eq.(\ref{CKMfit}) are obtained
under the assumption of the unitarity of the $3 \times 3$ CKM matrix. Giving
up this assumption the elements of the third row of the CKM matrix can deviate
substantially from the values in Eq.(\ref{CKMfit}).
\\
As an illustration we show the results of an analysis of the SM4, where
it is assumed that the four dimensional CKM matrix is unitary, while the 
three dimensional matrix does not have to be unitary \cite{Eberhardt:2010bm}, 
similar results were obtained in 
\cite{Bobrowski:2009ng,Das:2010fh,Alok:2010zj,Nandi:2010zx,Buras:2010pi}.
\\
In Fig. (\ref{Vtd},\ref{Vts},\ref{Vtb}) we show the possible values of $V_{td}$, $V_{ts}$ and
$V_{tb}$ in a complex plane and we compare it with the value from
Eq.(\ref{CKMfit}). The possible values of $V_{td}$, $V_{ts}$ and
$V_{tb}$ were obtained by replacing the unitarity
of the $3x3$ CKM matrix by the unitarity of the $4x4$ CKM matrix and by
demanding that
all direct measurements for the 
CKM elements, as well as bounds from FCNC and electro-weak 
precision observables are ful-filled.

For the last row of the CKM matrix still deviations of the order of $100\%$ 
(for $V_{td}$ and $V_{ts}$) from the
standard values in Eq.(\ref{CKMfit}) are possible. 
Here more precise determinations of $V_{tb}$ (single-top, $R$-ratio) and any idea how to 
determine $V_{td}$ and $V_{ts}$ directly (see e.g. \cite{Ali:2010xx}) would be extremly helpful.

\subsection{Another success of the CKM paradigm}

Another success of the CKM picture represents the rare penguin decay
$b \to s \gamma$ \cite{bsgamma}.
Experiment \cite{Asner:2010qj} agrees well with the NNLO theory prediction 
\cite{Misiak:2006zs}
\begin{eqnarray}
Br (b \to s \gamma)^{\rm Exp} & = & \left( 3.55 \pm 0.26 \right) \cdot 10^{-4} \; ,
\\
Br (b \to s \gamma)^{\rm Theo} & = & \left( 3.15 \pm 0.23 \right) \cdot 10^{-4} \; .
\end{eqnarray}
The experimental average uses numbers from
BaBar, BELLE and CLEO.
For a more comprehensive list of references see e.g.
\cite{Misiak:2010zz}.

\section{Hints for deviations from the SM}
Besides the impressive success of the CKM picture, we see currently
several hints for deviations of experiment from theory.

\subsection{$B_s$-mixing}

In the $B_s$ system the dimuon asymmetry measured by the D0 collaboration
\cite{Abazov:2010hv}, see also \cite{Williams:2011nc}, is a factor of 42 larger 
than the standard model prediction \cite{Lenz:2006hd}.
\begin{eqnarray}
A_{sl}^b & := & (0.506 \pm 0.043) a_{sl}^d + (0.494 \pm 0.043) a_{sl}^s \; ,
\\
A_{sl}^{b, \rm SM} & = & (-0.023+0.005-0.006)\%  \; ,
\\
A_{sl}^{b, \rm Exp} & = & (-0.957\pm0.251\pm0.146)\%  \; ,
\label{dimuonexp}
\end{eqnarray}
where $a_{sl}^q$ is the semileptonic CP asymmetry in the $B_q$-system, see e.g.
\cite{asl}. The SM prediction has been obtained by using the NLO-QCD calculations in
\cite{NLO}, see also \cite{NLOrome}.
The statistical significance of the deviation is $3.2 \; \sigma$\footnote{
After the conference D0 updated \cite{Abazov:2011yk} the measurement. 
First the coeffcients defining the dimoun asymmetry were updated:
$A_{sl}^b = (0.594 \pm 0.022) a_{sl}^d + (0.406 \pm 0.022) a_{sl}^s$.
Using also updated theory predictions for the mixing observables \cite{Lenz:2011ti}
one gets $A_{sl}^{b, \rm SM} =  (-0.023 \pm 0.004)\%$.
For the new measurement instead of $6.1$ fb$^{-1}$ now $9.0$ fb$^{-1}$ 
of data were used. As expected \cite{Lenz:2011zz} the central value went down, but the 
statistical significance increased to $3.9 \; \sigma$:
$A_{sl}^{b, \rm Exp} = (-0.787 \pm 0.172 \pm 0.093) \%$.},
which triggered a lot of interest. At the time of writing this proceedings the first
paper of \cite{Abazov:2010hv} had approximately 150 citations within 15 months.
\\
In terms of box diagrams the semileptonic CP-asymmetries are given by
      \begin{eqnarray}
      a_{sl}^q  :=  \frac{|\Gamma_{12}^q|}{|M_{12}^q|} \sin \left( \phi_q \right)
      &&
      \phi_q  :=  \arg \left( - \frac{M_{12}^q}{\Gamma_{12}^q}\right) \; .
      \end{eqnarray}
$M_{12}$ is expected to be very sensitive to new physics, while $\Gamma_{12}$ should
be free of new physics contributions within the hadronic uncertrainities (see
e.g. \cite{Lenz:2011zz}).
Therefore one can write in the presence of new physics model independently
      \begin{eqnarray}
      M_{12}^q = M_{12}^{q, \rm SM} \cdot { |\Delta_q| e^{i \phi_q^\Delta}}
      &&
      \Gamma_{12}^q = \Gamma_{12}^{q, \rm SM} \; .
      \label{NPinmix}
      \end{eqnarray}
Large new physics contributions to $\Gamma_{12}$ would also affect the lifetime
ratios of heavy hadrons - which agree relatively well, although there are
large uncertainties due to the non-perturbative bag parameters, see e.g.
\cite{Lenz:2011ti,Lenz:2008xt}) - and also the average number of charm quarks per
$b$-decays $n_c$ and $BR(b \to s + \mbox{no charm})$, see e.g. \cite{Lenz:2000kv}
for a mini-review.
\\
Now one can write the general expression for semi leptonic CP asymmetries in the presence
of new physics.
      \begin{eqnarray}
      a_{sl}^q =  \frac{|\Gamma_{12}^{q, \rm SM}|}{|M_{12}^{q, \rm SM}|} 
      \frac{\sin \left( \phi_q^{\rm SM} { + \phi_q^\Delta} \right)}{{ |\Delta_q|}}
\; .
      \label{NP-LN}
      \end{eqnarray} 
Using the SM predictions for the mixing parameters
\cite{Lenz:2006hd,Lenz:2011ti}
and the bounds on $\Delta_q$ from
\cite{Lenz:2010gu}
we get a maximal value of the dimuon asymmetry of
\begin{eqnarray}
A_{sl}^b  & = & (0.594 \pm 0.022) (5.4 \pm 1.0) 10^{-3} \frac{\sin (\phi_d^{\rm SM}+\phi_d^\Delta)}{|\Delta_d|} 
   \nonumber \\
          & + & (0.406 \pm 0.022) (5.0 \pm 1.1) 10^{-3} \frac{\sin (\phi_s^{\rm SM}+\phi_s^\Delta)}{|\Delta_s|} \; ,
\\
A_{sl}^b& \leq & -4.8 \cdot 10^{-3} \; ,
\label{1sigma}
\\
A_{sl}^b& \leq & -9.0 \cdot 10^{-3} \; .
\label{3sigma}
\end{eqnarray}
For the bound in Eq.(\ref{1sigma}) we have taken the $1 \sigma$ deviation
 of the SM predictions as well as the $1 \sigma$ deviation of the fit result for $\Delta_q$:
$(\phi_d^\Delta = -12.9^\circ-2.7^\circ; \phi_s^\Delta = - 51.6^\circ-9.7^\circ; 
               |\Delta_d| = 0.747-0.0079; |\Delta_s| = 0.887-0.064)$.
In Eq.(\ref{3sigma}) we show for comparison the 3 $\sigma$ deviations in all parameters
$(\phi_d^\Delta = -12.9^\circ-7.0^\circ; \phi_s^\Delta = - 90^\circ; 
               |\Delta_d| = 0.747-0.17; |\Delta_s| = 0.887-0.12)$.
\\
The measured value of the dimuon asymmetry in Eq.(\ref{dimuonexp}) is about 1.66 $\sigma$ 
above the bound in Eq.(\ref{1sigma}).
There are now 3 possibilities to explain this minor discrepancy:
\begin{enumerate}
\item The theory prediction for $\Gamma_{12}$ has considerably larger uncertainties, 
      than stated e.g. in
      \cite{Lenz:2011zz}.
      For this reasoning several counter-arguments can be given:
      \begin{itemize}
      \item The theoretical determination of lifetimes of heavy mesons relies 
            on the same footing as the determination of $\Gamma_{12}$, 
            see e.g. \cite{Lenz:2008xt}.
            Within the hadronic uncertainties epxeriments and theory predictions
            agree well. However, this comparison is affected by sizeable
            uncertainties of the non-perturbative bag parameters of four-quark operators,
            that appear in the HQE.
            Here clearly a theoretical (= lattice) improvement would be very helpful,
            see e.g. \cite{Lenz:2011zz} for more details.
      \item The higher order terms in the HQE converge well:\\
            The current status of the theory is discussed in \cite{Lenz:2011zz}.
            There it was shown, that the corrections to the leading term in the
            Heavy-Quark-Expansion (HQE) (i.e. $\alpha_s$, subleading $1/m_b$ and 
            lattice corrections to the vacuum insertion approximation for the non-perturbative
            matrix elements of the arising four-quark operators) give absolutely no hint for
            a non-convergence of the HQE. Their size is between $6 \%$ and $19 \%$. 
            This situation improved considerably compared to e.g. \cite{Lenz:2004nx}.
     \item  Another possibility to check the robustness of the theory prediction
            for $\Gamma_{12}$ is to use the exclusive approach for its determination.
            In the pioneering work of Aleksan et al. \cite{Aleksan:1993qp} this 
            complementary approach gave 
            \begin{eqnarray}
            \frac{\Delta \Gamma_s}{\Gamma_s} & \approx & {\cal O} (0.15) \; ,
            \end{eqnarray}
            which is in perfect agreement with the HQE determination \cite{Lenz:2011ti}
            \begin{eqnarray}
            \frac{\Delta \Gamma_s}{\Gamma_s} & \approx & 0.137 \pm  0.027\; .
            \end{eqnarray}
            There was a recent update of the work of Aleksan et al. in \cite{Chua:2011er}, 
            which again gives results that are in perfect agreement 
            with the HQE determination of $\Delta \Gamma_s$.
      \end{itemize}
      Besides these strong arguments for the validity of the HQE approach for $\Gamma_{12}$,
      one also has to keep in mind, how large the effects on $\Gamma_{12}$ would have to be, 
      to explain the above 1.66 $\sigma$ discrepancy:
      \begin{itemize}
      \item Assuming that there is new physics in $M_{12}$ and using for the size of the new 
            physics effects the fit results 
            from \cite{Lenz:2010gu}
            one would need an enhancement of $|\Gamma_{12}|$ of about $200 \%$
            to obtain the central value of the dimuon-asymmetry.
      \item Assuming that there is no new physics in $M_{12}$ and all the discrepancy is due
            to a failure of the HQE one would need an enhancement of $|\Gamma_{12}|$ of 
            about $4100 \%$ ($3300\%$ for the updated measurement of the dimuon asymmetry from D0)
            to obtain the central value of the dimuon-asymmetry.
       \end{itemize}
       So it seems very unplausible, that a failure of the HQE is responsible for the
       minor discrepancy in the dimuon asymmetry. Nethertheless
       a  measurement of $\Delta \Gamma_s$ in the near future at LHCb will be very important to
       settle this issue.
\item New physics contribution to $\Gamma_{12}$.
      \\
      Due to the arguments given below Eq. (\ref{NP-LN}) we consider it also impossible that
      new physics can give contributions of the order of $200 \%$ - $3300\%$.
      New physics effects of the order of the hadronic uncertainties are probably not yet 
      excluded.
\item The $1.66 \sigma$ discrepancy is just a statistical fluctuation.
      \\
      This seems currently to be by far the most obvious explanation. Actually
      the new D0 value for the dimuon asymmetry \cite{Abazov:2011yk} shrank - 
      the discrepancy is now 1.5 $\sigma$.
      \\
      To clarify my point of view: I do not consider the effect itself seen at D0 
      to be a statistical fluctuation, only the high central value. Even if the actual
      dimuon asymmetry is below the bound in Eq.(\ref{1sigma}) this would correspond 
      to a very large new physics effect.
      As will be discussed below there are more indications for new physics acting in 
      the $B_s$-system, that are consistent with a large dimoun asymmetry 
      (and also with the sign).
      Of course this point of view will change if the central value stays, 
      when the errors are reduced. But compared to 1.5 standard deviations 
      I consider the above
      arguments for a validity of the theory approach to be much stronger. 
\end{enumerate}
Another hot topic in the $B_s$-mixing system is the angular analysis of 
the decay $B_s \to J/\psi \phi$.
This decay is investigated at TeVatron (CDF and D0) \cite{Abbott:2011hj}
and LHCb \cite{angular}.
There are also some small hints for deviations from the SM, which are compatible with the
sign and size of the dimuon asymmetry.
From the angular analysis one gets $\Delta \Gamma_s$ and $S_{\psi \phi}$, which
is defined as
      \begin{equation}
      S_{\psi \phi}^{\rm SM} = \sin \left(2 \beta_s { - \phi_s^\Delta} 
                           - \delta_s^{\rm Peng, SM}
                           - { \delta_s^{\rm Peng, NP}} \right) \; .
      \end{equation}
Neglecting penguins, one gets in the standard model
      \begin{equation}
      S_{\psi \phi}^{\rm SM} = 0.0036 \pm 0.002 \; . 
      \end{equation}
The new physics fit from \cite{Lenz:2010gu} gave however
      \begin{equation}
      S_{\psi \phi} = 0.78^{+0.12}_{-0.19}\; .
      \end{equation}
The history of these measurements, which deviated originally even more from the SM prediction
is presented in \cite{Abbott:2011hj}.
After the conference D0 updated its analysis with 8.0 fb$^{-1}$ of data \cite{D0update},
with the result
\begin{eqnarray}
\phi_s^{J/\psi \phi} & =: & - 2 \beta_s { + \phi_s^\Delta} 
                            + \delta_s^{\rm Peng, SM}
                            + { \delta_s^{\rm Peng, NP}}
\\                   & =  & -0.55^{+0.38}_{-0.36} \; ,
\\
\Delta \Gamma_s & = & 0.163^{+0.065}_{-0.064} \; \mbox{ps}^{-1} \; .
\end{eqnarray}
Here we will soon get a definite answer from LHCb, whether a large NP contribution exists in
$ S_{\psi \phi}$ or not.
The LHCb status with the 2010 data \cite{LHCb} is given in the following figure.
\begin{center}
\includegraphics[width=0.9 \textwidth,angle=0]{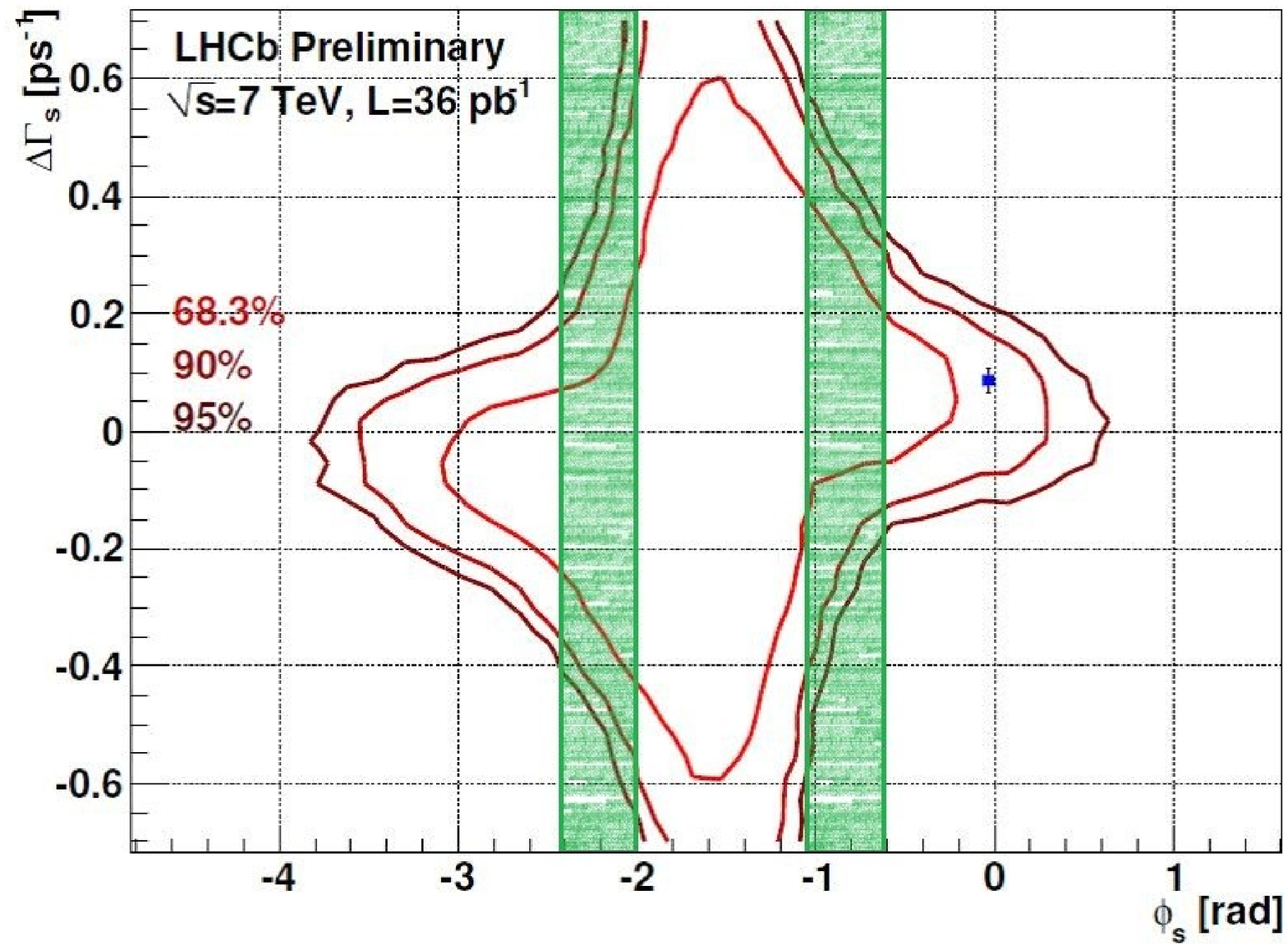}
\end{center}

\vspace{-4cm}

The blue point denotes the SM value.
In green the fit result for the new physics phase $\phi_s^\Delta$ from \cite{Lenz:2010gu} 
is shown under the assumption that penguins
are negligible. The coming result from LHCb will probably have about 10 times more statistics.
\\
Finally we would like to make a comment on the often used relation 
\begin{eqnarray}
a_{sl}^s & = & - \frac{\Delta \Gamma}{\Delta M} 
\frac{S_{\psi \phi}}{\sqrt{1- S_{\psi \phi^2}}} { \cdot \delta} 
\label{relation} \; ,
\end{eqnarray}
with
\begin{eqnarray}
\delta & = & \frac{\tan \left({    \phi_s^{\rm SM} }+ \phi_s^\Delta \right)}
                  {\tan \left({ -2 \beta_s^{\rm SM}}+ \phi_s^\Delta + \delta_s^{\rm peng, SM}+ \delta_s^{\rm peng, NP} \right)} \; .
\end{eqnarray}
Typically it is assumed that for a large new physics phase in mixing $\phi_s^\Delta$,
$\delta$ is closed to one
and can be neglected. A more detailed analysis in \cite{Lenz:2011zz} shows however, that
$\delta = 1$ is strongly violated. Therefore Eq.(\ref{relation}) without $\delta$  
can not be used to eliminate  the theory prediction for $\Gamma_{12}$. 
Instead it might be used to determine the size
of penguin contributions to the decay $B_s \to J/\psi \phi$, which is also an important task.

\subsection{$B_d$-mixing}

Due to the increased precision in experiment and theory in the last years hints for new
physics were also found in the golden plated mode 
$B_s \to J/\psi K_s$ \cite{Bigi:1981qs}.
Comparing the direct measurement \cite{Sahoo:2011kf} of $\sin (2 \beta)$ 
with the indirect determination from the CKM fits \cite{Lenz:2010gu} we get
\begin{eqnarray}
\sin \left( 2 \beta \right)^{\rm Exp.} & < &
\sin \left( 2 \beta \right)^{\rm Fit } \; \; \; ,
\\
0.678 \pm 0.020  &<& 0.831^{+0.013}_{-0.030} \; \; \; ,
\\
\beta = \left( 21.4 \pm 0.8        \right)^\circ &<& 
        \left(28.09^{+0.7}_{-1.49} \right)^\circ \; \; \; .
\end{eqnarray}
This discrepancy was pointed out first in \cite{Lunghi:2008aa}
and then in \cite{Buras:2008nn}, it is currently seen by all CKM-fitting groups,
with a similar statistical significance:
\begin{displaymath}
\begin{array}{|l|l|l|}
\hline
\mbox{Reference}                         & \mbox{Group}                       & \mbox{Deviation}
\\
\hline
\hline
\mbox{1102.3917} \; \cite{Laiho:2011nz}  & \mbox{Laiho, Lunghi, Van de Water} & 2.5 - 3.3 \; \sigma
\\
\hline
\mbox{1010.6069} \; \cite{Lunghi:2010gv} & \mbox{Lunghi, Soni}                & 3.3 \; \sigma
\\
\hline
\mbox{1010.5089} \; \cite{Bevan:2010gi}  & \mbox{UTfit}                       & 2.6 \; \sigma
\\
\hline
\mbox{1008.1593} \; \cite{Lenz:2010gu}   & \mbox{Lenz, Nierste, CKMfitter}    & 2.8 \; \sigma
\\
\hline
\end{array}
\end{displaymath}

\subsection{More hints for deviations}

\subsubsection{$ B \to \tau \nu$}

The discrepancy in $ B \to \tau \nu$ was already mentioned in the 
discussion of $V_{ub}$. The measured branching ratio \cite{btaunu} is
considerably larger than the theoretically expected one.
If this tension is due to new physics, it might 
be triggered by direct contributions to the decay $ B \to \tau \nu$ (e.g.
Two-Higgs-Doublett model) or by new physics in $B_d$ mixing, which results in
a different value for $V_{ub}$.

\subsubsection{$\epsilon_K$}

The discrepancies in the CKM fits mentioned above in the $B_d$-section might also
be due to new physics acting in $\epsilon_K$ \cite{Buras:2008nn}.
$\epsilon_k$ depends strongly on $V_{cb}$ (fourth power) 
as well as on the value the non-perturbative bag-paramter
$\hat{B}_K$. 
Depending on the central values and errors used for $\hat{B}_K$ one gets different 
discrepancies between experiment and standard model prediction for $\epsilon_K$, e.g.
\begin{displaymath}
\begin{array}{|l|l|c|c|}
\hline
\mbox{Reference}                        & \mbox{Group}                      & \hat{B}_K             & \mbox{Deviation of} \; \; \epsilon_K
\\
\hline
\hline
\mbox{1102.3917} \; \cite{Laiho:2011nz} & \mbox{Laiho, Lunghi, Van de Water} & 0.736 \pm 0.020 & 1.9 \; \sigma
\\
\hline
\mbox{1008.1593} \; \cite{Lenz:2010gu}  & \mbox{Lenz, Nierste, CKMfitter}    &0.724 \pm 0.067 & 0.5 \; \sigma 
\\
\hline
\end{array}
\end{displaymath}
A nice comparison of the different methods used by the different groups to determine lattice averages 
is given in \cite{Wingate:2011fb}.

\subsubsection{$B \to K^{(*)}ll$}

The status of the decays $B \to K^{(*)}ll$ was discussed in \cite{gudrun}. There were 
some hints for discrepancies
between the data from BaBar, Belle and CDF and the standard model expectation. After 
the Conference
LHCb \cite{LHCbKstar} and CDF \cite{:2011ja} announced more precise measurements of 
$B \to K^{(*)}ll$, which are consistent with
the standard model predictions, see e.g. \cite{SMKll}.
Here still more data are needed to draw some final conclusions.

\subsubsection{$ B_s \to \mu \mu$}
The very rare decay $ B_s \to \mu \mu$ gives strong contraints on many extensions of the
standard model. The experimental situation was discussed in \cite{Btomumu-TeVatron} for the TeVatron
and in \cite{Btomumu-LHC} for LHC.
After the conference CDF announced a first two-sided bound on
the rare decay $B_s \to \mu   \mu$ \cite{Collaboration:2011fi}
\begin{equation}
Br(B_s \to \mu   \mu) = \left( 18^{+11}_{-9} \right) \cdot 10^{-9}
\end{equation}
Updating the SM prediction (see e.g. \cite{Buchalla:1995vs})
we obtain with the input parameters from \cite{Lenz:2010gu}
\begin{equation}
Br(B_s \to \mu   \mu) = \left( 3.0 \pm 0.4 \right)  \cdot 10^{-9} \; .
\end{equation}
If we in addition assume that there is now new physics in $\Delta M_s$
we can get rid of the large uncertainties due to the decay constant and
get the very precise prediciton
\begin{equation}
Br(B_s \to \mu   \mu) = \left( 3.1 \pm 0.1 \right)  \cdot 10^{-9} \; .
\end{equation}
The central value of the experimental number from CDF 
is a factor of 6 larger than the theory
prediction, but the statistical significance of the deviation is less than
2 $\sigma$.
LHCb \cite{LHCbmumu} and CMS \cite{Collaboration:2011kr}
presented new results for $B_s \to \mu \mu$ at the EPS conference.
They did not confirm the signal of CDF, but there is still a lot of room for new
physics.
\begin{eqnarray}
\mbox{CMS:} \; \; Br(B_s \to \mu   \mu) & < & 19 \cdot 10^{-9}   \; \; \; 95\% C.L. \; ,
\\
\mbox{LHCb:} \; \; Br(B_s \to \mu   \mu) & < & 15 \cdot 10^{-9}   \; \; \; 95\% C.L. \; .
\end{eqnarray}
For the $B_d$-decay the following bounds hold now
\begin{eqnarray}
\mbox{CMS:} \; \; Br(B_d \to \mu   \mu) & < & 4.6 \cdot 10^{-9}   \; \; \; 95\% C.L. \; ,
\\
\mbox{LHCb:} \; \; Br(B_d \to \mu   \mu) & < & 5.2 \cdot 10^{-9}   \; \; \; 95\% C.L. \; .
\end{eqnarray}
\subsubsection{Hadronic B decays}

Hadronic decays were discussed in \cite{hadroni}.

\subsection{Combining the hints}

In \cite{Lenz:2010gu} a model independent fit of new physics acting only in the
neutral meson mixing systems was performed.
Besides the usual parameters, now also the parameters $\Delta_d$ and $\Delta_s$ defined
in Eq.(\ref{NPinmix}) were fitted. The results are shown in the following two figures:
\begin{center}
\includegraphics[width=0.49\textwidth,angle=0]{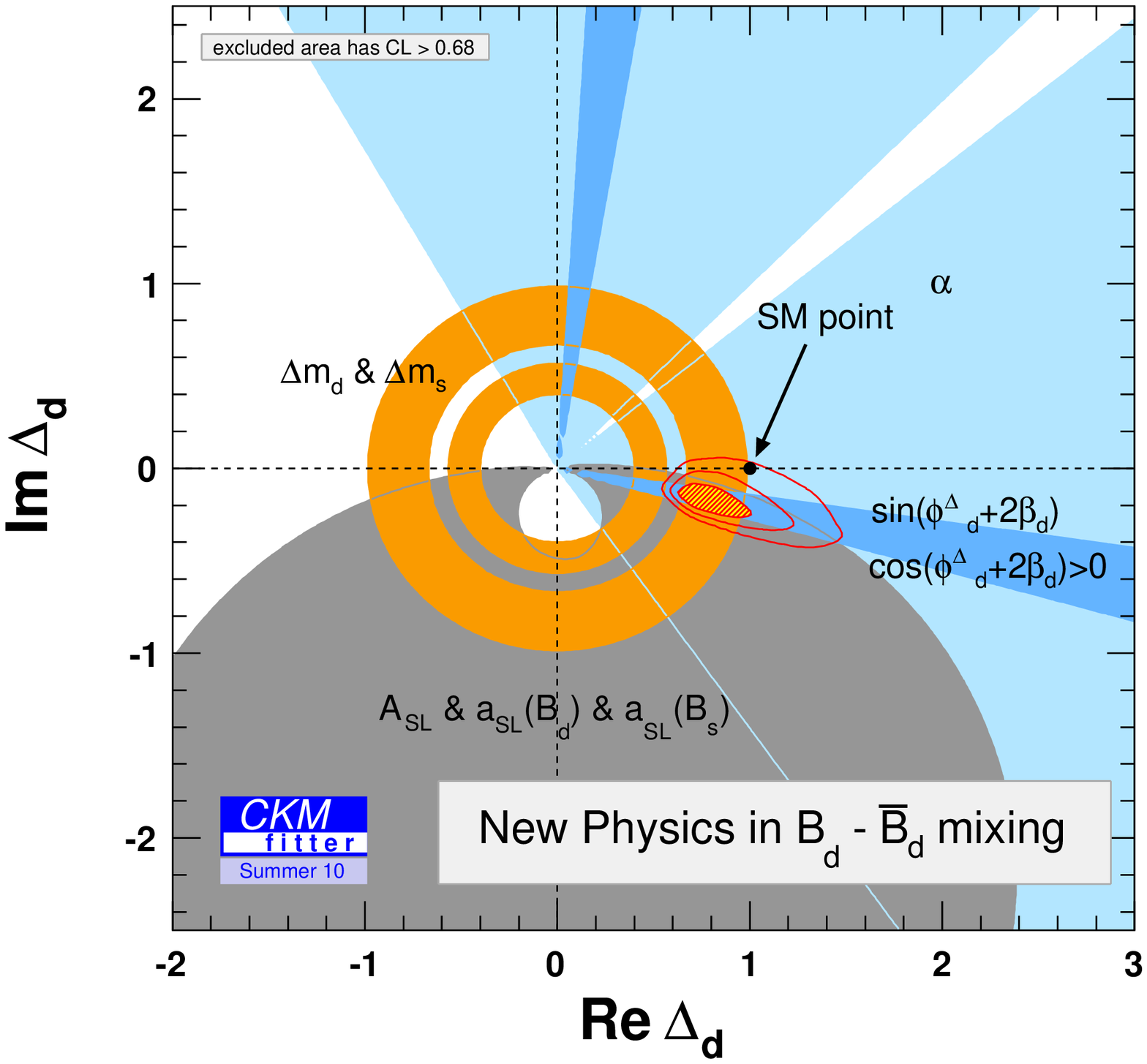}
\hfill
\includegraphics[width=0.49\textwidth,angle=0]{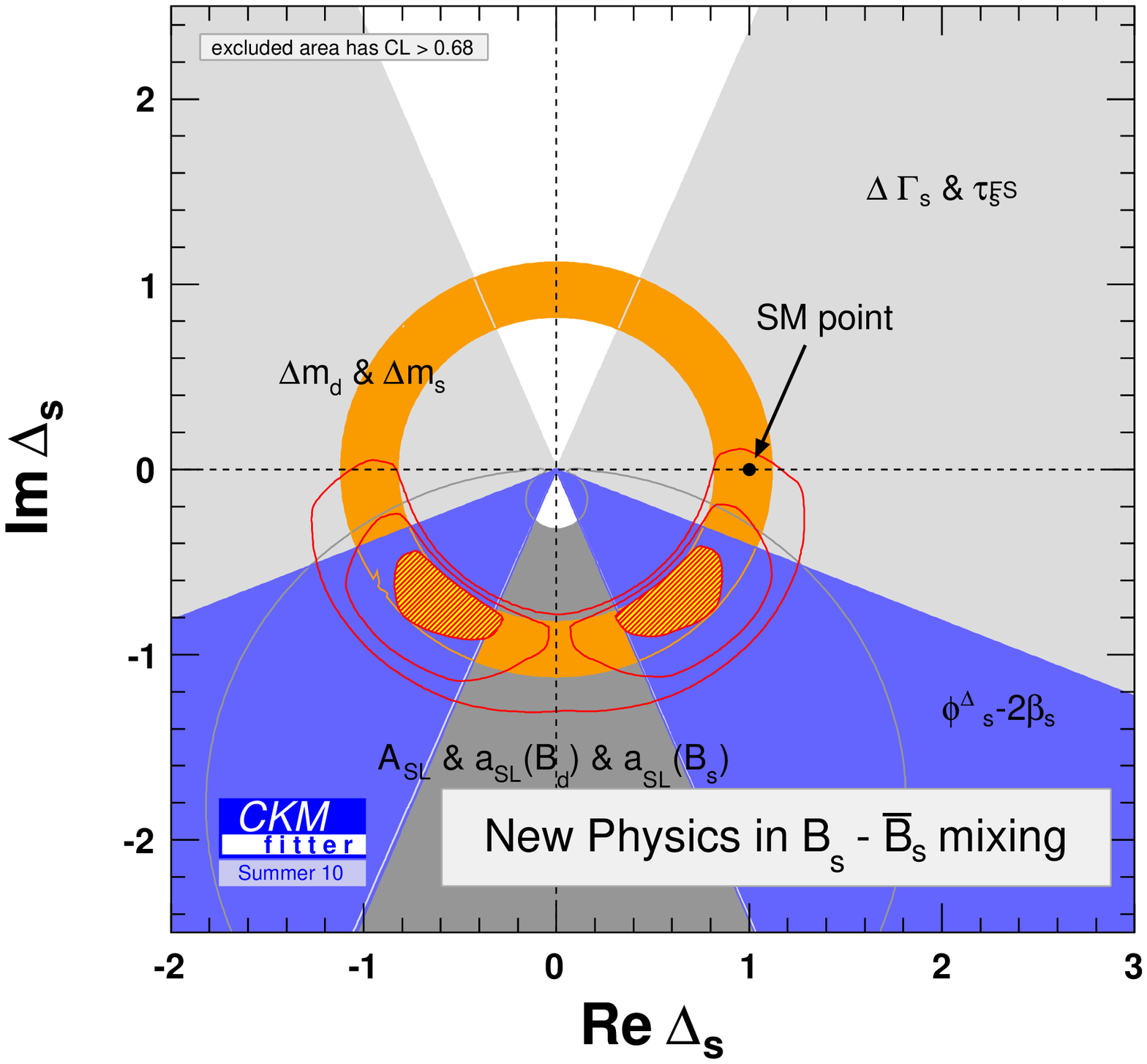}
\end{center}
The standard model corresponds to $\Delta_q = 1$. The fit shows however that
Im $\Delta_d$ = 0 = Im $\Delta_s$  is excluded with 3.8 standard deviations.
Including the the new D0 dimuon result, the SM would probably be 
excluded with more than 4 standard deviations.

\section{Conclusion and Outlook}

Soon we will get much more data, in particular from the LHC, but also from Tevatron 
and the B-factories.
A first part of this new data was already made public after FPCP2011.
\\ 
We will know definitely, whether $S_{\psi \phi}$ is really large
and  whether the current hints for large new physics effects in
$B$-mixing are real.
We also will soon know how large $\Delta \Gamma_s$ is. This measurement
will have a big impact on our understanding of the theory methods used 
to determine $\Gamma_{12}$.
Moreover we will gain more information on the semileptonic CP asymmetries.
Tevatron might still improve on the dimuon asymmetry $A_{sl}^b$, the B-factories, Tevatron and LHC might
enhance our knowledge about the individual semi leptonic CP asymmetries $a_{sl}^{s,d}$
and LHCb will have information on $a_{sl}^{s} - a_{sl}^{d}$.
The data and bounds on the rare decays $B \to K^{(*)}ll$ and $B_s \to \mu \mu$
will be improved. Even if the parameter space for new physics effects was shrinking
recently, there is still room for large new effects.
Tevatron and LHC will also provide more data for charm mixing \cite{charm}. Here it is still
not clear how well our theoretical tools from the beauty sector work
\cite{kagan}. The $D$-system is also well suited to search for new physics effects, see e.g.
\cite{Golowich:2007ka}.
Also rare Kaon decays like $K \to \pi \nu \nu$ \cite{kaon}
or lepton flavor violating decays like $\mu \to e \gamma$
\cite{meg} will play a crucial role.
\\
Besides identifying the new physics effects in certain observables or in a combined
fit, the next important question will be
{\it How to interpret this data?} \cite{NPinter}.
This can be done within certain models for extensions of the standard model (see
the plethora of papers on the arxive) or model independently.
Since the CKM picture works very well, the framework of minimal flavor violation (MFV) 
\cite{MFV} seems to be very promising. Nethertheless one should keep in mind
that there are also viable counter examples for MFV like the SM4: due to possible 
cancellations between $ \delta V_{td,s,b}$ and the $t'$-loop, there can still
be effects of ${\cal O} (100 \%)$ in $B$-mixing, which are consistent with the CKM
fits, see e.g. 
\cite{Eberhardt:2010bm,Bobrowski:2009ng,Das:2010fh,Alok:2010zj,Nandi:2010zx,Buras:2010pi}.
A very promising approach to interpet the hints for new physics is to study 
correlations among different obsverables in 
different models, as strongly advocated by the group of A. Buras.
\\
In order to make full use of the coming data, there is however also
a lot of  basic work (i.e. no model buildung...) to be done. We finish therefore
with a {\it What to do list}.
\section{What to do list}
We have to improve our current knowledge of the CKM matrix:
\begin{itemize}
\item First row: Understand the origin of the different results for $V_{ub}$ and improve the 
      accuracy in $V_{us}$.
\item Second row: Improve the accuracy in $V_{cs}$ and $V_{cb}$.
\item Third row: Improve the accuracy in $V_{tb}$ and find a way to measure $V_{td}, V_{ts}$.
\end{itemize}
We also have to improve our understanding of the machinery to describe the mixing systems
theoretically:
\begin{itemize}
\item Test of the HQE with lifetimes of heavy hadrons.
      \begin{itemize}
      \item $\tau_{B^+}/ \tau_{B_d}$ and $\tau_{B_s}/ \tau_{B_d}$ fit well within the
            hadronic uncertainites and we
            have currently no hints for deviations from the HQE.
      \item To improve, precise non-perturbative matrix elements for the arising 
            4-quark operators are urgently needed.
      \item There are also some perturbative improvements of lifetime predictions missing, like
            the full NLO-QCD calculation of the $\Lambda_b$ lifetime.
      \end{itemize}
\item More precise theoretical predictions for mixing observables.
      \begin{itemize}
      \item Precise decay constants and Bag parameter for $\Delta M$
      \item Additional Bag parameters at dimension 6 and 7 for $\Gamma_{12}$
      \item $\alpha_s/m_b$ corrections for $\Gamma_{12}$
      \item $\alpha_s^2$ corrections for $\Gamma_{12}$
      \end{itemize}
\item Theoretical predictions for charm mixing observables
      \\
      Push HQE to its limits.
\item Try to improve the exclusive approach as a cross-check, see e.g.
      \cite{Chua:2011er}.
\end{itemize}
Finishing these tasks will enable us to make full use of the
exciting times lying ahead of us.

\Acknowledgements
I would like to thank the organizers of FPCP2011 for the invitation and
for their great work in creating such a pleasant und fruitful atmossphere 
during the conference.

\end{document}